\documentclass[11pt, preprint]{article}
\usepackage{color} \usepackage{cancel}
\usepackage{float} \usepackage{graphicx}% Include figure files
\usepackage{subfigure} \usepackage{caption}	% Advanced figure caption
\usepackage{dcolumn}% Align table columns on decimal point
\usepackage[toc,page]{appendix} \usepackage{authblk}	 		% Title page
\usepackage{indentfirst}	% Indent for first paragraph
\usepackage{authblk}	 		% Title page
\usepackage{multirow}     % Multi-row cell in table
\usepackage{hyperref}     % Internal and external hyper links
\usepackage{amsfonts, amssymb, amsmath}
\usepackage[left=2.5cm,right=2.5cm,top=2.9cm,bottom=2.9cm,a4paper]{geometry}
\usepackage[noadjust]{cite}
\usepackage{filecontents}

\usepackage{soul}
\usepackage{xcolor}
\setstcolor{red}

\graphicspath{{./figures/}}
\hypersetup{ colorlinks   = true, % Color links instead of ugly boxes
urlcolor     = blue, % Color of external hyperlinks
linkcolor    = blue, % Color of internal links
citecolor    = red 	 % Color of citations }
}

\title{\Large\bf A new approach to exact solutions construction in scalar cosmology with a Gauss-Bonnet term}
\author[1]{I. V. Fomin\thanks{ingvor@inbox.ru}}
\author[2,3]{S. V. Chervon\thanks{chervon.sergey@gmail.com}}
\affil[1]{\small \it  Bauman Moskow State Technical University, 2-nd Baumanskaya street, 5, Moscow, 105005, Russia}
\affil[2]{\small \it Ulyanovsk State Pedagogical University, Ulyanovsk, 100-years V.I. Lenin's Birthday Square, B.4, 432071, Russia}
\affil[3]{\small \it Astrophysics and Cosmology Research Unit, School of Mathematics, Statistics and Computer Science, University of KwaZulu-Natal, Private Bag X54 001, Durban 4000, South Africa}

\begin{document}

\maketitle
%----------------------------------------------------------------
\begin{abstract}
We study the cosmological model based on Einstein-Gauss-Bonnet gravity with non-minimal coupling of a scalar field to a Gauss-Bonnet term in 4D Friedmann universe.  We show how constructing the exact solutions by the method based on a confrontation of the Hubble parameter in the model under consideration with that in a standard scalar field inflationary cosmology.
\end{abstract}
%\tableofcontents
%----------------------Introduction------------------------------
\section{Introduction}

The cosmological model based on Einstein-Gauss-Bonnet (EGB) gravity with non-minimal coupling of a scalar field to a Gauss-Bonnet (GB) term in 4D Friedmann universe is of growing interest. The reasons are as follow.

Although the scenarios of cosmological inflation~\cite{Starobinsky:1980te,Guth:1980zm,Albrecht:1982wi,Linde:1981mu,Mukhanov:1990me,Olive:1990sb,Martin:2013tda} explain the origin of a large-scale structure and correspond to observational data~\cite{Hinshaw:2012aka,Ade:2015xua,Ade:2015lrj}, there are problems that go beyond the standard inflationary scenario, for example, quantum gravity.
Approaching to the very early universe one can consider Einstein gravity with some corrections as an effective theory of the quantum gravity. The effective supergravity action from superstrings induces correction terms of higher order in the curvature and they may play a significant role in the early Universe. One of such simple correction is the GB term~\cite{Zwiebach:1985uq,Zumino:1985dp}.

It is well known, that the variation of the GB term is different from zero only in a space with dimensions of not less than five. Evidently, the four-dimensional space does not satisfy this condition, but if the non-minimal coupling of a scalar field with the GB term is taken into consideration, the dynamical equations of cosmology are quite different from the standard inflationary cosmology~\cite{Guo:2010jr,Koh:2016abf}.
Then, influence of the GB term in 4D universe is effective.

The observational constraints on slow-roll inflation in the models with the scalar fields coupled to the GB term, and the exact and approximate expressions for parameters of cosmological perturbations were obtained in the papers~\cite{Guo:2010jr,Koh:2016abf,Koh:2014bka,vandeBruck:2015gjd}.

The models of dark energy with the GB term, in the context of second accelerated expansion of the universe~\cite{Perlmutter:1998np,Riess:1998cb} were considered in the works~\cite{Nojiri:2005vv,Carter:2005fu,Leith:2007bu};
EGB gravity was used in the context of  reconstruction of gravity theories from the universe expansion history~\cite{Cognola:2006sp,Nojiri:2010wj} as well.

In the paper~\cite{Fomin:2017vae}, we paid attention for two main problems: the estimation of the influence of the non-minimal coupling of the scalar field with the GB term on the cosmological dynamics and finding the effective technique of exact solutions constructions.
Both these problems were successfully solved after finding the connection between the Hubble parameters $\overline{H}$ and $H$ for standard scalar field cosmology and for scalar cosmology, based on EGB gravity with non-minimal coupling to a scalar field,  correspondingly. Both of the models are studied for a spatially-flat Friedmann universe.

In the present work we generalize the proposed method to the case of an open and closed Friedmann universe, besides new solutions for a spatially-flat universe are presented as well. It should be noted here, that we used the presentation of open and closed universes as a spatially-flat one filled by a perfect fluid \cite{Barrow:2016qkh}.

The article is organized as follow.
In Sec.\ref{section2} we consider the dynamical equations of non-minimal coupled scalar field with the GB term in Friedmann universe.
In Sec.\ref{Conformity} the conformity between 4D EGB cosmological models and standard Friedmann cosmology is considered.

In Sec.\ref{examples} we obtain the exact solutions for the models with De Sitter and power-law expansion by
choosing the evolutionary laws for the scale factor $a(t)$ and the scalar field $\phi(t)$. We were also convinced that with a special choice of the model's parameters, the Hubble parameters $H$ and $\overline{H}$ are coincided.

In Sec.\ref{SEcH} the exact solutions for the power-law scale factor were obtained by the setting of the relation
between $H$ and $\overline{H}$. Also, we consider the another ansatz, $\overline{H}=0$, leading to the models without a connection with standard cosmology and we obtain the solutions for $k\neq0$ and the general solutions for $k=0$.

In Sec.\ref{GW-Standard} we consider the representation of time derivative of the GB coupling function $\xi$, which allows us to compare not only the Hubble parameters, but also other parameters of the standard cosmological inflationary model and models with a non-minimal coupling of the scalar field with the GB term.

In the Sec.\ref{Conclusion} we summarize the results of the work .

\section{The model's dynamical equations}\label{section2}

The action under consideration is composed of the Einstein-Hilbert term and a canonical scalar field non-minimally interacting to the GB term through a coupling function $\xi(\phi)$~\cite{Guo:2010jr,Koh:2016abf}
\begin{equation}
\label{action}
S=\int d^4x\sqrt{-g}\left[\frac{1}{2} R - \frac{1}{2}g^{\mu\nu}\partial_{\mu}\phi \partial_{\nu}\phi
 - V(\phi)-\frac12\xi(\phi) R_{GB}^2\right]
\end{equation}
where $R^{2}_{GB} = R_{\mu\nu\rho\sigma} R^{\mu\nu\rho\sigma}- 4 R_{\mu\nu} R^{\mu\nu} + R^2$ is the GB term. The GB coupling function $\xi(\phi)$ is required to be a function of the scalar field in order to give nontrivial effects on the cosmological dynamics. To describe a homogeneous and isotropic universe we chose the Friedmann-Robertson-Walker (FRW) metric in the form
\begin{equation}
ds^2=-dt^2+a^2(t)\left(\frac{d r^2}{1-k r^2}+r^2 d\Omega^2\right)
\end{equation}
where $a(t)$ is a scale factor, a constant  $k$ is the indicator of universe's
type:\footnote{It is usually accepted do not normalise $k$ to the unity ($k=+1,0,-1$) in
the models where  the curvature term $k/a^2$ is considered as the term appeared in a spatially-flat
universe from a perfect fluid \cite{Barrow:2016qkh}. }
$ k>0,~k=0,~k<0 $ are associated with closed, spatially-flat, open universes, correspondingly.

The dynamical equations of cosmology for the model (\ref{action}) are~\cite{Koh:2016abf}
\begin{eqnarray} \label{beq2a}
&& 3H^2 =\frac{1}{2}\dot{\phi}^2 + V(\phi) -\frac{3k}{a^2}+ 12\dot{\xi}H\left(H^2+\frac{k}{a^2}\right)\\
\label{beq3a}
&& \dot{H} = -\frac{1}{2}\dot{\phi}^2+\frac{k}{a^2} +2\ddot{\xi}\left(H^2+\frac{k}{a^2}\right)+2\dot{\xi}H\left(2\dot{H}-H^2-\frac{3k}{a^2}\right)\\
&& \ddot{\phi} + 3 H \dot{\phi} + V_{,\phi} +12 \xi_{,\phi} \left(H^2+\frac{k}{a^2}\right) \left(\dot{H}+H^2\right) = 0 \label{beq4a}
\end{eqnarray}
where a dot represents a derivative with respect to the cosmic time $t$,
$H \equiv \dot{a}/a$ denotes the Hubble parameter, $V_{,\phi} = \partial V/\partial \phi$, and $\xi_{,\phi} = \partial \xi/\partial \phi$. Since $\xi$ is a function of $\phi$, $\dot{\xi}$ implies $\dot{\xi} = \xi_{\phi} \dot{\phi}$. If $\xi$ is a constant,
the dynamical equations would not be influenced by the GB term in four dimensional spacetime.

The equation (\ref{beq4a}) can be derived from the equations (\ref{beq2a})--(\ref{beq3a}). Therefore, we will consider these equations as fully describe the cosmological dynamics.

Let us mention about one interesting interpretation of an open and closed universes as a spatially-flat one filled by
the
perfect fluid with the relation $3k=\rho_{m0}$ ~\cite{Barrow:2016qkh}. In such approach, we have the inflationary stage which is driven by the scalar field in the spatially-flat Friedmann universe ($k=0$).

If $\xi=const$, the equations (\ref{beq2a})--(\ref{beq3a}) are reduced to the equations for standard scalar field inflation
\begin{eqnarray}
\label{st1}
&& 3H^2 =\frac{1}{2}\dot{\phi}^2 + V(\phi) -\frac{3k}{a^2}\\
\label{st2}
&& \dot{H} = -\frac{1}{2}\dot{\phi}^2+\frac{k}{a^2}
\end{eqnarray}
%%%
In the next section we will introduce new notation for the Hubble parameter $H$ with the aim to differ it from the same in the 4D EGB cosmological model (\ref{action}).

\section{Conformity between standard Friedmann and 4D EGB cosmology}\label{Conformity}
To make possible a generation of exact solutions for inflation in 4D EGB cosmology and comparison with standard Friedmann cosmology let us define the new (generalized, in respect to that introduced in \cite{Fomin:2017vae}) connection between the Hubble parameter $\overline{H}$ of standard inflation and the Hubble parameter $H$ in the 4D EGB model %with GB term
\begin{equation}
\label{connection}
\overline{H}=H-2\dot{\xi}\left(H^{2}+\frac{k}{a^{2}}\right)
\end{equation}
%%%
The equations (\ref{beq2a})--(\ref{beq3a}), in this case, can be rewritten as
\begin{equation}
\label{ex1}
\frac{1}{2}\dot{\phi}^{2}+V(\phi)=-3H^{2}+6\overline{H}H+\frac{3k}{a^{2}}
\end{equation}
\begin{equation}
\label{ex2}
\frac{1}{2}\dot{\phi}^{2}=-\dot{\overline{H}}+\overline{H}H-H^{2}+\frac{k}{a^{2}}
\end{equation}
$\xi=const$ implies $H=\overline{H}$ than the equations (\ref{ex1})--(\ref{ex2})
are reduced to (\ref{st1})--(\ref{st2}).

Further, after simple transformation, we can rewrite the equations (\ref{connection})--(\ref{ex2}) in the following form
\begin{equation}
\label{ex3}
V(\phi)=-2H^{2}+5\overline{H}H+\dot{\overline{H}}+\frac{2k}{a^{2}}
\end{equation}
\begin{equation}
\label{ex4}
\frac{1}{2}\dot{\phi}^{2}=-\dot{\overline{H}}+\overline{H}H-H^{2}+\frac{k}{a^{2}}
\end{equation}
\begin{equation}
\label{exi}
\dot{\xi}=\frac{H-\overline{H}}{2\left(H^{2}+\frac{k}{a^{2}}\right)}
\end{equation}
%%%
Since the equations (\ref{ex3})--(\ref{exi}) contain five unknown functions, to generate the exact solutions without additional conditions it is necessary to set two of them.

\section{The exact solutions from given evolutions of scale factor and scalar field}\label{examples}
In the papers~\cite{Ellis:1990wsa,Chervon:1997yz} the exact solutions for standard cosmology were obtained by the choice of evolutionary laws
for the scalar field $\phi=\phi(t)$ or for the scale factor $a=a(t)$.
Now, we will generate the exact solutions by the choice of both: the evolutionary law for the scalar field $\phi=\phi(t)$ and for the scale factor $a=a(t)$. Then we compare the dynamics for 4D EGB and standard Friedmann cosmology by means of Hubble parameters.

\subsection{De Sitter expansion}
For example, consider the cosmological model with De Sitter expansion
\begin{eqnarray}
\label{a1}
&&a(t)=a_{0}\exp(At), \,\,\, H=A\\
\label{phi1}
&&\phi(t)=Bt
\end{eqnarray}
%%%
where $A(>0)$ and $B$ are arbitrary constants.

Form equations (\ref{ex3})--(\ref{exi}) we obtain
\begin{equation}
\overline{H}(t)=A+\frac{B^{2}}{2A}-\frac{k}{3Aa^{2}_{0}}e^{-2At}
\end{equation}
\begin{equation}
\xi(t)=-\left(\frac{1}{12A^{4}}+\frac{B^{2}}{8A^{4}}\right)\ln\left|A^{2}a^{2}_{0}+ke^{-2At}\right|
-\frac{B^{2}}{4A^{3}}t+const
\end{equation}
\begin{equation}
\xi(\phi)=-\left(\frac{1}{12A^{4}}+\frac{B^{2}}{8A^{4}}\right)\ln\left|A^{2}a^{2}_{0}+ke^{-\frac{2A}{B}\phi}\right|
-\frac{B}{4A^{3}}\phi+const
\end{equation}
\begin{equation}
V(\phi)=3A^{2}+\frac{5}{2}B^{2}+\frac{k}{a^{2}_{0}}e^{-\frac{2A}{B}\phi}
\end{equation}
%%%
For a spatially-flat Friedmann universe ($k=0$) we have
\begin{equation}
\overline{H}(t)=A+\frac{B^{2}}{2A}
\end{equation}
\begin{equation}
\xi(\phi)=-\left(\frac{1}{12A^{4}}+\frac{B^{2}}{8A^{4}}\right)\ln A^{2}a^{2}_{0}-\frac{B}{4A^{3}}\phi+const
\end{equation}
\begin{equation}
V(\phi)=3A^{2}+\frac{5}{2}B^{2}
\end{equation}
%%%
If $B=0$, we have $H=\overline{H}=A$, $\xi=const$, $V=V_{E}=3A^{2}$ and $\phi=\phi_{E}=0$, where $\phi_{E}$ and
$V_{E}$ are the scalar field and the potential for standard inflation.

\subsection{Power-law evolution of the scale factor}
Now, we consider a power-law scale factor evolution and a logarithmic evolution of the scalar field: % with
\begin{eqnarray}
\label{apl1}
&&a(t)=a_{0}t^{m}, \,\,\, H=m/t\\
\label{phipl1}
&&\phi(t)=C\ln(Bt),
\end{eqnarray}
%%%
where $m(>0)$, $B(>0)$ and $C$ are arbitrary constants.

The coupling function $\xi$ for an arbitrary $m$ is defined from (\ref{exi}) in quadratures only. Nevertheless,
 an explicit dependence $\xi=\xi(t)$ and, respectively, $\xi=\xi(\phi)$ can be found for the specific values of $m$.
Explicit integration, for example, can be performed for the model with $m=2$.

Setting $m=2$, form equations (\ref{ex3})--(\ref{exi}) we obtain
\begin{equation}
\overline{H}(t)=\frac{C^{2}}{6t}+\frac{4}{3t}-\frac{k}{5a^{2}_{0}t^3}
\end{equation}
\begin{equation}
\xi(t)=\left(-\frac{C^{2}}{96}+\frac{1}{24}\right)B^{2}t^{2}+\left(\frac{kC^{2}}{384a^{2}_{0}}
+\frac{k}{480a^{2}_{0}}\right)
\ln\left|4a^{2}_{0}t^{2}+k\right|+const
\end{equation}
\begin{equation}
\xi(\phi)=\left(-\frac{C^{2}}{96}+\frac{1}{24}\right)e^{\frac{2\phi}{C}}+\left(\frac{kC^{2}}{384a^{2}_{0}}
+\frac{k}{480a^{2}_{0}}\right)
\ln\left|\frac{4a^{2}_{0}}{B^{2}}e^{\frac{2\phi}{C}}+k\right|+const
\end{equation}
\begin{equation}
V(\phi)=\frac{B^{2}}{10a^{2}_{0}}\left(15C^{2}a^{2}_{0}e^{\frac{2\phi}{C}}+40a^{2}_{0}e^{\frac{2\phi}{C}}
+6kB^{2}\right)e^{-\frac{4\phi}{C}}
\end{equation}
%%%
For a spatially-flat Friedmann universe ($k=0$) we have the following solution:
\begin{equation}
\overline{H}(t)=\frac{C^{2}}{6t}+\frac{4}{3t}
\end{equation}
\begin{equation}
\xi(\phi)=\left(-\frac{C^{2}}{96}+\frac{1}{24}\right)e^{\frac{2\phi}{C}}+const
\end{equation}
\begin{equation}
V(\phi)=\frac{B^{2}}{10}\left(15C^{2}+40\right)e^{-\frac{2\phi}{C}}
\end{equation}
%%%
In the case of $C=\pm2$, we have the solution: $H=\overline{H}=2/t$, $\phi(t)=\phi_{E}(t)=\pm2\ln(Bt)$, $\xi=const$ and $V(\phi)=V_{E}(\phi)=10B^{2}e^{\mp2\phi}$.

It should be note here, that one can construct new exact solutions from the equations (\ref{ex3})--(\ref{exi}) using another value of $m$ for the power-law scale factor (\ref{apl1}).

\section{The exact solutions from ansatzs}\label{SEcH}
In Sec.\ref{examples} we are setting the evolutionary laws for a scale factor and a scalar field with the aim to solve 4D EGB cosmological dynamics equations (\ref{ex3})--(\ref{exi}).
There is one another way to solve these equations. Namely, we can impose the ansatz on the Hubble parameters $F(H,\overline{H})=0$ and make a suitable choice of one of them.
This will correctly define the system of equations (\ref{ex3})--(\ref{exi}) and we have to find three unknown functions: $ \overline{H}$ (from the ansatz $F(H,\overline{H})=0$, if $H=H(t)$ is given), $\phi(t),~\xi(\phi)$.

Now, we consider the application of the special ansatzs for constructing the exact solutions.

\subsection{The first ansatz}
The first ansatz is
\begin{equation}
\label{anz1}
-\dot{\overline{H}}+\overline{H}H-H^{2}=0
\end{equation}
%%%
When the scale factor and the Hubble parameter are as in the case of power-law universe expansion (\ref{apl1}), taking into account the case of $k\neq0$, from ansatz (\ref{anz1}) we obtain
\begin{equation}
\label{anz1H}
\overline{H}(t)=\frac{m^{2}}{(1+m)t}
\end{equation}
%%%
The Hubble parameters $H$ and $\overline{H}$ can be equal when $m=0$ or $a=a_{0}$ i.e. at the beginning of inflation, only.

As in Sec.\ref{examples}, we can find the explicit dependence $\xi=\xi(t)$ for the specific values of $m$.

Setting $m=2$ we have
\begin{equation}
\phi(t)=\frac{\sqrt{2k}}{a_{0}t}
\end{equation}
\begin{equation}
\xi(t)=\frac{1}{24}t^{2}-\frac{k}{96a^{2}_{0}}\ln\left|4a^{2}_{0}t^{2}+k\right|+const
\end{equation}
\begin{equation}
\xi(\phi)=\frac{k}{12a^{2}_{0}\phi^{2}}-\frac{k}{96a^{2}_{0}}\ln\left|\frac{8k}{\phi^{2}}+k\right|+const
\end{equation}
\begin{equation}
V(\phi)=\frac{a^{2}_{0}}{6k}\phi^{2}(3\phi^{2}+20)
\end{equation}
%%%
It is worth to note, that in the case of open Friedmann universe ($k=-1$) we have the phantom scalar
field with negative kinetic energy.

\subsection{The second ansatz}
In the context of the second ansatz, we consider the specific model without connection with standard cosmology.
This is the case when $\overline{H}=0$.
With this suggestion the equations (\ref{ex3})--(\ref{exi}) are reduced to the following form
\begin{equation}
\label{ex3zero}
V(\phi)=-2H^{2}+\frac{2k}{a^{2}}
\end{equation}
\begin{equation}
\label{ex4zero}
\frac{1}{2}\dot{\phi}^{2}=-H^{2}+\frac{k}{a^{2}}
\end{equation}
\begin{equation}
\label{exizero}
\dot{\xi}=\frac{H}{2\left(H^{2}+\frac{k}{a^{2}}\right)}
\end{equation}
%%%
Now, we can rewrite the equation (\ref{ex4zero}) in terms of $a(t)$ from the definition of $H=\dot{a}/a$ as
\begin{equation}
\label{phizero}
\dot{a}^{2}+\frac{1}{2}(a\dot{\phi})^{2}=k
\end{equation}
%%%
Firstly, we consider the model with $k=1$, $k>0$ and scale factor
\begin{equation}
a(t)=A\sin(Bt),
\end{equation}
%%%
with additional condition $k=A^{2}B^{2}$, where $A$ and $B$ are positive constants.

From equations (\ref{phizero}), (\ref{ex3zero}) and (\ref{exizero}) we obtain
\begin{equation}
\phi(t)=\pm\sqrt{2}Bt, \,\,\,\,    V=2B^{2}
\end{equation}
\begin{equation}
\xi(t)=-\frac{1}{4B^{2}}\ln\left|\cos(2Bt)+3\right|+const
\end{equation}
\begin{equation}
\xi(\phi)=-\frac{1}{4B^{2}}\ln\left|\cos(\sqrt{2}\phi)+3\right|+const
\end{equation}
%%%
For the open Friedmann universe ($k=-1$) we consider the phantom scalar field satisfying the equation
\begin{equation}
\label{phizerophantom}
\dot{a}^{2}-\frac{1}{2}(a\dot{\phi})^{2}=-1
\end{equation}
%%%
Here, in equation (\ref{phizerophantom}), we changed the sign before kinetic energy $\dot{\phi}^{2}/2$ in equation (\ref{phizero}) for the case of the phantom field.

The next example of the exact solution is given for us by setting the scale factor as
\begin{equation}
a(t)=A\cosh(Bt)
\end{equation}
%%%
with the restriction $A^{2}B^{2}=1$. Thus, we have the solution
\begin{equation}
\phi(t)=\pm\sqrt{2}Bt, \,\,\,\,    V=2B^{2}
\end{equation}
\begin{equation}
\xi(t)=\frac{1}{4B^{2}}\ln\left|\cosh(2Bt)-3\right|+const
\end{equation}
\begin{equation}
\xi(\phi)=\frac{1}{4B^{2}}\ln\left|\cosh\left(\sqrt{2}\phi\right)-3\right|+const
\end{equation}
%%%
For a spatially-flat Friedmann universe ($k=0$) we have the general solution of equations (\ref{ex3zero})--(\ref{exizero}) with the phantom field
\begin{equation}
\label{gen}
\phi(t)=\pm\sqrt{2}\ln(a(t))+c
\end{equation}
\begin{equation}
\label{gen1}
V(t)=-2\left(\frac{\dot{a}}{a}\right)^{2}
\end{equation}
\begin{equation}
\label{gen2}
\xi(t)=\frac{1}{2}\int\frac{a}{\dot{a}}dt
\end{equation}
%%%
where $c$ is a constant of integration.

As the new example, we give the solution with the scale factor
\begin{equation}
a(t)=A\exp(Bt^{m})
\end{equation}
The Hubble parameter is $H=mBt^{m-1}$.
From equations (\ref{gen})--(\ref{gen2}) we obtain
\begin{eqnarray}
&&\phi(t)=\pm\sqrt{2}Bt^{m}+c_{1}, \,\,\,\,\, c_{1}=c\pm\sqrt{2}\ln A\\
&&V(\phi)=-2B^{2}m^{2}\left(\pm\frac{\phi-c_{1}}{\sqrt{2}B}\right)^{\frac{2(m-1)}{m}}\\
&&\xi(t)=\frac{t^{2-m}}{2(2-m)mB}+const\\
&&\xi(\phi)=\frac{1}{2(2-m)mB}\left(\pm\frac{\phi-c_{1}}{\sqrt{2}B}\right)^{\frac{2-m}{m}}+const
\end{eqnarray}
%%%
For the special cases $m=1/2$ and $m=1/3$ we have the inverse potentials $V(\phi)\propto-(\phi-c_{1})^{-2}$ and $V(\phi)\propto-(\phi-c_{1})^{-4}$.

\section{Generation of 4D EGB exact solutions from a scale factor}\label{GW-Standard}

Now, we consider  the following representation of time derivative of the GB coupling function (ansatz)
\begin{equation}
\label{xik}
\dot{\xi}=\frac{Ca^{3}}{\dot{a}^{2}+k}
\end{equation}
%%%
where $C$ is an arbitrary constant. In the paper~\cite{Fomin:2017vae} we considered the case where $C=1/2$, $k=0$.

With the representation (\ref{xik}) the equations  (\ref{beq2a})--(\ref{beq3a}) are reduced to
the ones for the standard-like cosmology (\ref{st1})--(\ref{st2}). Discarding the Hubble parameter via  a scale factor we transform these equations to the following form
\begin{eqnarray}
\label{Vk}
&&V(\phi)=\frac{\ddot{a}}{a}+2\frac{\dot{a}^{2}}{a^{2}}+\frac{2k}{a^{2}}-12C\dot{a}\\
\label{phik}
&&\frac{1}{2}\dot{\phi}^{2}=-\dot{H}+\frac{k}{a^{2}}
\end{eqnarray}
%%%
The difference from standard cosmology is the last term in the potential, therefore we can represent it as $V=V_{E}+V_{GB}$.
The first term $V_{E}$ is the potential in standard Friedman cosmology and the second term $V_{GB}=-12C\dot{a}$ corresponds to the GB correction. When $C=0$ one has $V_{GB}=0$ and $\xi=const$.

From equation (\ref{connection}) with the ansatz on the coupling function (\ref{xik}) we obtain the Hubble parameter $\overline{H}$ for the model with minimal coupling %(\ref{action})
\begin{equation}
\label{HubbleGB}
\overline{H}(t)=H(t)-2Ca(t)
\end{equation}
%%%
The influence of non-minimal coupling of a scalar field to the GB term on the cosmological dynamics is defined by the sign of $C$. When $C>0$ the non-minimal coupling rises the expansion of the universe in respect to standard Friedmann cosmological model with minimal coupling and it
 downwards the universe expansion in the case of $C<0$. For $C=0$ we have the standard Friedmann cosmology with $\xi=const$ and $H=\overline{H}$.

Also, from the relation
\begin{equation}
\frac{\ddot{a}}{a}=H^{2}+\dot{H}
\end{equation}
%%%
we obtain the connection between accelerations of the universe in the cases of standard and 4D EGB cosmology
\begin{equation}
\frac{\ddot{\overline{a}}}{\overline{a}}=\frac{\ddot{a}}{a}-2Ca(3H-2Ca)
\end{equation}

Thus, the difference between dynamics in standard and 4D EGB cosmology doesn't depend on a value of $k$ in the framework of chosen ansatz (\ref{xik}).

Now, we consider the model with oscillating scale factor, besides $k\neq0$.
\begin{equation}
\label{scalefactor}
a(t)=A\cos(Bt)
\end{equation}
%%%
where $A$ and $B$ are positive constants.

From equations (\ref{xik})--(\ref{phik}) we obtain
\begin{equation}
\label{potentialt}
V(t)=2B^{2}\tan^{2}(Bt)-B^{2}+\frac{2k}{A^{2}\cos^{2}(Bt)}+12ABC\sin(Bt)
\end{equation}
\begin{equation}
\label{field}
\phi(t)=\pm\frac{\sqrt{2A^{2}B^{2}+2k}}{AB}\ln\left|\sec(Bt)+\tan(Bt)\right|+c_{1}
\end{equation}
\begin{equation}
\label{couplingfunctiont}
\xi(t)=-\frac{AC}{B^{3}}\sin(Bt)+\left(\frac{A^{2}C}{B^{2}\sqrt{k}}+\frac{C\sqrt{k}}{B^{4}}\right)
\arctan\left[\frac{AB}{\sqrt{k}}\sin(Bt)\right]+const
\end{equation}
%%%
where $c_{1}$ is a constant of integration.

The Hubble parameter is
\begin{equation}
\label{HubbleGB-2}
H(t)=-B\tan(Bt)
\end{equation}
%%%
To simplify the formulae, we introduce two new functions
\begin{equation}
\label{Functionf}
f(\phi)=\exp\left(\pm\frac{2AB(\phi-c_{1})}{\sqrt{2A^{2}B^{2}+2k}}\right)
\end{equation}
\begin{equation}
\label{Functionf}
g(\phi)=\frac{\sqrt{f^{2}(\phi)+1}}{f(\phi)+1}
\end{equation}
%%%
From the expression (\ref{field}) we have
\begin{equation}
\label{time}
Bt=\arctan\left(\frac{f(\phi)-1}{f(\phi)+1}\right)
\end{equation}
%%%
One can easily obtains the dependence $\xi=\xi(\phi)$ by means of the substitution (\ref{time}) into (\ref{couplingfunctiont})
\begin{eqnarray}
\label{couplingfunctionphi}
\xi(\phi)=-\frac{AC}{B^{3}}\sin\left(\arctan\left(\frac{f(\phi)-1}{f(\phi)+1}\right)\right)+\nonumber
\left(\frac{A^{2}C}{B^{2}\sqrt{k}}+\frac{C\sqrt{k}}{B^{4}}\right)\times\\
\times\arctan\left[\frac{AB}{\sqrt{k}}\sin\left(\arctan\left(\frac{f(\phi)-1}{f(\phi)+1}\right)\right)\right]+const
\end{eqnarray}
%%%
Now, we can find the potential as the function of a scalar field
\begin{eqnarray}
\label{potentialphi}
&&V(\phi)=\frac{1}{A^{2}(f(\phi)+1)^{2}g(\phi)}\Big[(6\sqrt{2}A^{3}BC+A^{2}B^{2})f^{2}(\phi)+\\ \nonumber
&&+(4kf^{2}(\phi)+A^{2}B^{2}+4k)g(\phi)-6A^{2}B^{2}f(\phi)g(\phi)-6\sqrt{2}A^{3}BC\Big]
\end{eqnarray}
%%%
Further, we redefined the scalar field as $\varphi=\phi-c_{1}+30$.

The corresponding potential $V=V(\varphi)$ with
specific values of model's parameters is represented in Figure $1$.

\begin{figure}[ht]
\begin{center}
{\includegraphics*[scale=1]{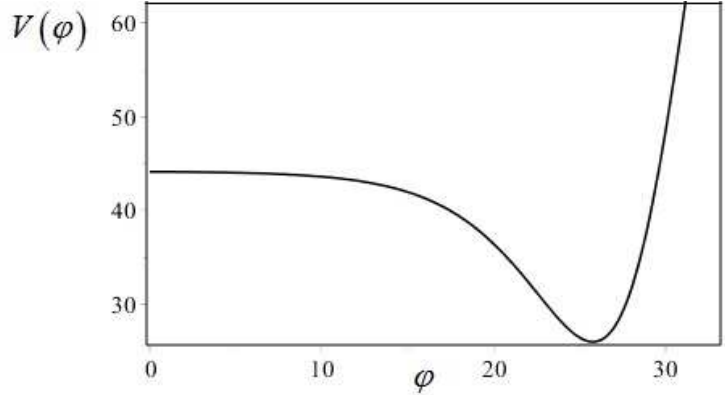}}
\end{center}
\caption{Potential $V=V(\varphi)$ with parameters $k=30$, $A=1.2$, $B=1$, $C=6$. Parameters $A$ and $C$ regulate the height of the potential at the point $V(\varphi=0)$. Parameter $B$ regulates the slope of the potential from $V(\varphi=0)$ to $V_{min}$.
The shape of $V(\varphi)$ corresponds to Coleman-Weinberg potential~\cite{Olive:1990sb}.
}
\end{figure}
The potential of the similar form we obtain for inflation in the case of closed ($k=1$) Friedmann universe,  but for open ($k=-1$)  universe with $V(\varphi)>0$ one must include the positive cosmological constant $\Lambda$ in the model. For this aim, one can simply redefine the potential as $V(\varphi)\rightarrow V(\varphi)+\Lambda$.

Let us note here, that
it is possible to obtain the similar potential $V_{E}(\varphi)$ for standard cosmology ($C=0$, $k\neq0$) with another model's parameters. This fact stress that we found new physically important exact solutions with the potential of Coleman-Weinberg type for 4D EGB and standard cosmology using the proposed method.

\section{Conclusion and discussions}\label{Conclusion}

In the present work we continue to develop the methods of exact solutions construction in 4D EGB cosmology with a non-minimal coupling of the scalar field to the GB term. The main idea of generation the exact solutions from standard Friedmann cosmology was proposed in our earlier work \cite{Chervon:1997yz}. Here we extended the results for
such cosmological models with non-zero index of curvature $k$ or with perfect fluid of the special kind.

The key role of generating new exact solutions is contained in the formula (\ref{connection}) where the conformity between standard Friedmann and 4D EGB cosmology is setting up. Such connection gave possibility to transform the model's dynamical equations to the form (\ref{ex3})-(\ref{exi}) which was used in two directions: the setting of a scale factor and evolution of the scalar field (Sec.\ref{examples}) or imposing  the ansatz on the Hubble parameters of standard Friedmann $\overline{H}$  and 4D EGB cosmology $H:  F(H,\overline{H})=0$ (Sec.\ref{SEcH}).

New set of exact solutions was generated using the representation of time derivative of the GB coupling function in the form (\ref{xik}). This leads to the connection between Hubble parameters (\ref{HubbleGB}) which allowed to generate new exact solutions for the giving scale factor $a(t)$. For the solution (\ref{HubbleGB-2}) we obtained the potential of the form similar to Coleman-Weinberg one \cite{Olive:1990sb}.

In the case of a spatially-flat Friedmann universe we can estimate the influence of the GB term on the dynamics by means of difference between $e$-folds numbers for 4D EGB and standard inflation
\begin{equation*}
\Delta_{N}=N-\overline{N}=\int_{t_{i}}^{t_{e}}(H-\overline{H})dt=2\int_{t_{i}}^{t_{e}}\dot{\xi}H^{2}dt,
\end{equation*}
%%%
where $t_{i}$ and $t_{i}$  are the times of the beginning and the end of inflation.

For $\xi=const$ we obtain $\Delta_{N}=0$, in the case of $\overline{H}=0$ we have $\dot{\xi}=1/2H$ and
\begin{equation*}
N=\int_{t_{i}}^{t_{e}}Hdt
\end{equation*}
%%%
For models with coupling function (\ref{xik}) we have
\begin{equation*}
\Delta_{N}=N-\overline{N}=2C\int_{t_{i}}^{t_{e}}a(t)dt,
\end{equation*}
%%%
thus, difference between $e$-folds numbers depends on the dynamics and the value of constant $C$.

Let us note also, that it is possible to generate the exact solutions for 4D EGB cosmology on the basis of exact solutions for standard cosmology in Friedmann universe for the cases $k=0$ and $k\neq0$ from the equations (\ref{xik})--(\ref{phik}). This procedure for a spatially-flat Friedmann universe was proposed in the paper~\cite{Fomin:2017vae}, the examples of exact solutions for open and closed Friedmann models one can find, for instance, in~\cite{Ellis:1990wsa,Chervon:1997yz,Barrow:2016qkh}.

\section{Acknowledgements}

This work was finalized during the authors' visit in 2017 to the University of KwaZulu-Natal. They are thankful to ACRU and the Director Professor S.D. Maharaj for warm hospitality and the NFR for financial support. S.V. Chervon also acknowledges for financial support from the University of Zululand and Professor A. Beesham for kind attention.

I.V. Fomin was supported by RFBR grants 16-02-00488 A and 16-08-00618 A.

%%%%====================bibliography=================================


\begin{thebibliography}{99}

%\cite{Starobinsky:1980te}
\bibitem{Starobinsky:1980te}
  A.~A.~Starobinsky,
  ``A New Type of Isotropic Cosmological Models Without Singularity,''
  Phys.\ Lett.\  {\bf 91B} (1980) 99.
 % doi:10.1016/0370-2693(80)90670-X

%\cite{Guth:1980zm}
\bibitem{Guth:1980zm}
  A.~H.~Guth,
  ``The Inflationary Universe: A Possible Solution to the Horizon and Flatness Problems,''
  Phys.\ Rev.\ D {\bf 23}, 347 (1981).
  %doi:10.1103/PhysRevD.23.347

%\cite{Albrecht:1982wi}
\bibitem{Albrecht:1982wi}
  A.~Albrecht and P.~J.~Steinhardt,
  ``Cosmology for Grand Unified Theories with Radiatively Induced Symmetry Breaking,''
  Phys.\ Rev.\ Lett.\  {\bf 48}, 1220 (1982).
  %doi:10.1103/PhysRevLett.48.1220

%\cite{Linde:1981mu}
\bibitem{Linde:1981mu}
  A.~D.~Linde,
  ``A New Inflationary Universe Scenario: A Possible Solution of the Horizon, Flatness, Homogeneity, Isotropy and Primordial Monopole Problems,''
  Phys.\ Lett.\ B {\bf 108}, 389 (1982).
  %doi:10.1016/0370-2693(82)91219-9

%\cite{Mukhanov:1990me}
\bibitem{Mukhanov:1990me}
  V.~F.~Mukhanov, H.~A.~Feldman and R.~H.~Brandenberger,
  %``Theory of cosmological perturbations. Part 1. Classical perturbations. Part 2. Quantum theory of perturbations. Part 3. Extensions,''
  Phys.\ Rept.\  {\bf 215} (1992) 203.
  doi:10.1016/0370-1573(92)90044-Z

%\cite{Olive:1990sb}
\bibitem{Olive:1990sb}
  K.~A.~Olive,
  ``Lectures on particle physics and cosmology,''
  LAPP-TH-317-90, UMN-TH-910-90, C90-06-18.
  %%CITATION = LAPP-TH-317-90, UMN-TH-910-90, C90-06-18;%%

%\cite{Martin:2013tda}
\bibitem{Martin:2013tda}
  J.~Martin, C.~Ringeval and V.~Vennin,
  ``Encyclop�dia Inflationaris,''
  Phys.\ Dark Univ.\  {\bf 5-6}, 75 (2014).
  %doi:10.1016/j.dark.2014.01.003
  %[arXiv:1303.3787 [astro-ph.CO]].

%\cite{Hinshaw:2012aka}
\bibitem{Hinshaw:2012aka}
  G.~Hinshaw {\it et al.} [WMAP Collaboration],
  ``Nine-Year Wilkinson Microwave Anisotropy Probe (WMAP) Observations: Cosmological Parameter Results,''
  Astrophys.\ J.\ Suppl.\  {\bf 208}, 19 (2013);
  %doi:10.1088/0067-0049/208/2/19
  %[arXiv:1212.5226 [astro-ph.CO]];

\bibitem{Ade:2015xua}
  P.~A.~R.~Ade {\it et al.} [Planck Collaboration],
  ``Planck 2015 results. XIII. Cosmological parameters,''
  arXiv:1502.01589 [astro-ph.CO].

%\cite{Ade:2015lrj}
\bibitem{Ade:2015lrj}
  P.~A.~R.~Ade {\it et al.} [Planck Collaboration],
  ``Planck 2015 results. XX. Constraints on inflation,''
  arXiv:1502.02114 [astro-ph.CO].
  %%CITATION = ARXIV:1502.02114;%%
  %426 citations counted in INSPIRE as of 14 Dec 2015

%\cite{Zwiebach:1985uq}
\bibitem{Zwiebach:1985uq}
  B.~Zwiebach,
  ``Curvature Squared Terms and String Theories,''
  Phys.\ Lett.\  {\bf 156B} (1985) 315.
  % doi:10.1016/0370-2693(85)91616-8

%\cite{Zumino:1985dp}
\bibitem{Zumino:1985dp}
  B.~Zumino,
  ``Gravity Theories in More Than Four-Dimensions,''
  Phys.\ Rept.\  {\bf 137} (1986) 109.
%  doi:10.1016/0370-1573(86)90076-1

%\cite{Guo:2010jr}
\bibitem{Guo:2010jr}
  Z.~K.~Guo and D.~J.~Schwarz,
  ``Slow-roll inflation with a Gauss-Bonnet correction,''
  Phys.\ Rev.\ D {\bf 81}, 123520 (2010)
  %doi:10.1103/PhysRevD.81.123520
  %[arXiv:1001.1897 [hep-th]].

%\cite{Koh:2016abf}
\bibitem{Koh:2016abf}
  S.~Koh, B.~H.~Lee and G.~Tumurtushaa,
``Reconstruction of the Scalar Field Potential in Inflationary Models with a Gauss-Bonnet term,''
  arXiv:1610.04360 [gr-qc].

%\cite{Koh:2014bka}
\bibitem{Koh:2014bka}
  S.~Koh, B.~H.~Lee, W.~Lee and G.~Tumurtushaa,
  ``Observational constraints on slow-roll inflation coupled to a Gauss-Bonnet term,''
  Phys.\ Rev.\ D {\bf 90}, no. 6, 063527 (2014);
  %doi:10.1103/PhysRevD.90.063527
  [arXiv:1404.6096 [gr-qc]].

%\cite{vandeBruck:2015gjd}
\bibitem{vandeBruck:2015gjd}
  C.~van de Bruck and C.~Longden,
  ``Higgs Inflation with a Gauss-Bonnet term in the Jordan Frame,''
  Phys.\ Rev.\ D {\bf 93} (2016) no.6,  063519
 % doi:10.1103/PhysRevD.93.063519
  [arXiv:1512.04768 [hep-ph]].
  %%CITATION = doi:10.1103/PhysRevD.93.063519;%%

%\cite{Perlmutter:1998np}
\bibitem{Perlmutter:1998np}
  S.~Perlmutter {\it et al.} [Supernova Cosmology Project Collaboration],
  %``Measurements of Omega and Lambda from 42 high redshift supernovae,''
  Astrophys.\ J.\  {\bf 517} (1999) 565
 % doi:10.1086/307221
  [astro-ph/9812133].

%\cite{Riess:1998cb}
\bibitem{Riess:1998cb}
  A.~G.~Riess {\it et al.} [Supernova Search Team],
  ``Observational evidence from supernovae for an accelerating universe and a cosmological constant,''
  Astron.\ J.\  {\bf 116} (1998) 1009
%  doi:10.1086/300499
  [astro-ph/9805201].

%\cite{Nojiri:2005vv}
\bibitem{Nojiri:2005vv}
  S.~Nojiri, S.~D.~Odintsov and M.~Sasaki,
``Gauss-Bonnet dark energy,''
  Phys.\ Rev.\ D {\bf 71} (2005) 123509
 % doi:10.1103/PhysRevD.71.123509
  [hep-th/0504052].
  %%CITATION = doi:10.1103/PhysRevD.71.123509;%%

%\cite{Carter:2005fu}
\bibitem{Carter:2005fu}
  B.~M.~N.~Carter and I.~P.~Neupane,
``Towards inflation and dark energy cosmologies from modified Gauss-Bonnet theory,''
  JCAP {\bf 0606} (2006) 004
 % doi:10.1088/1475-7516/2006/06/004
  [hep-th/0512262].

%\cite{Leith:2007bu}
\bibitem{Leith:2007bu}
  B.~M.~Leith and I.~P.~Neupane,
  ``Gauss-Bonnet cosmologies: Crossing the phantom divide and the transition from matter dominance to dark energy,''
  JCAP {\bf 0705} (2007) 019
  %doi:10.1088/1475-7516/2007/05/019
  [hep-th/0702002].

%\cite{Cognola:2006sp}
\bibitem{Cognola:2006sp}
  G.~Cognola, E.~Elizalde, S.~Nojiri, S.~Odintsov and S.~Zerbini,
  ``String-inspired Gauss-Bonnet gravity reconstructed from the universe expansion history and yielding the transition from matter dominance to dark energy,''
  Phys.\ Rev.\ D {\bf 75} (2007) 086002
  %doi:10.1103/PhysRevD.75.086002
  [hep-th/0611198].

%\cite{Nojiri:2010wj}
\bibitem{Nojiri:2010wj}
  S.~Nojiri and S.~D.~Odintsov,
  ``Unified cosmic history in modified gravity: from F(R) theory to Lorentz non-invariant models,''
  Phys.\ Rept.\  {\bf 505} (2011) 59
  %doi:10.1016/j.physrep.2011.04.001
  [arXiv:1011.0544 [gr-qc]].

%\cite{Fomin:2017vae}
\bibitem{Fomin:2017vae}
  I.~V.~Fomin and S.~V.~Chervon,
  ``Exact Inflation in Einstein-Gauss-Bonnet Gravity,''
  arXiv:1704.03634 [gr-qc].

%\cite{Barrow:2016qkh}
\bibitem{Barrow:2016qkh}
  J.~D.~Barrow and A.~Paliathanasis,
  ``Observational Constraints on New Exact Inflationary Scalar-field Solutions,''
  Phys.\ Rev.\ D {\bf 94} (2016) no.8,  083518
 % doi:10.1103/PhysRevD.94.083518
  [arXiv:1609.01126 [gr-qc]].
  %%CITATION = doi:10.1103/PhysRevD.94.083518;%%

%\cite{Ellis:1990wsa}
\bibitem{Ellis:1990wsa}
  G.~F.~R.~Ellis and M.~S.~Madsen,
  ``Exact scalar field cosmologies,''
  Class.\ Quant.\ Grav.\  {\bf 8} (1991) 667.
 % doi:10.1088/0264-9381/8/4/012
  %%CITATION = doi:10.1088/0264-9381/8/4/012;%%

\bibitem{Chervon:1997yz}
  S.~V.~Chervon, V.~M.~Zhuravlev and V.~K.~Shchigolev,
  ``New exact solutions in standard inflationary models,''
  Phys.\ Lett.\ B {\bf 398} (1997) 269

%%\cite{Mathew:2016anx}
%\bibitem{Mathew:2016anx}
%  J.~Mathew and S.~Shankaranarayanan,
% ``Low scale Higgs inflation with Gauss–Bonnet coupling,''
%  Astropart.\ Phys.\  {\bf 84} (2016) 1
% % doi:10.1016/j.astropartphys.2016.07.004
%  [arXiv:1602.00411 [astro-ph.CO]].
%  %%CITATION = doi:10.1016/j.astropartphys.2016.07.004;%%
\end{thebibliography}
\end{document}